# Quantum rate theory of the trapping of hydrogen and deuterium by a vacancy in iron

I.H.Katzarov, A.T.Paxton

*Department of Physics, King's College London, Strand, London, WC2R 2LS, United Kingdom*

**Abstract**

We apply quantum rate theory to calculate the transition rates as hydrogen or deuterium atoms escape from a vacancy trap in iron into a neighbouring metastable site. We determine transition rates and corresponding activation energies over a wide range of temperatures covering both the quantum and classically dominated regimes. We find that quantum effects lead to an increase of the transition rate activation energy and to very significant recrossing of the transition state dividing surface. As a result of recrossing quantum transition state theory overestimates the rate of proton transfer by more than an order of magnitude and the rate of deuteron transfer by a factor of two.

**INTRODUCTION**

Hydrogen diffusion in Fe and Fe alloys is extremely important because it leads to engineering problems caused by hydrogen embrittlement and degradation of high-strength steels. A common factor in all H-assisted damage mechanisms is the crucial role of H transport and trapping. The high diffusivity of H in α-Fe (among the highest reported for any metal [1-4]) results from the very low activation energies due to the quantum nature of H [5]. The existence of microstructural imperfections (vacancies, solute atoms, dislocations, grain boundaries, etc.) introduces low energy trapping sites within the lattice which retard the overall diffusion rate [6-8]. Because H is a light element, intrinsic processes in H diffusion are strongly influenced by its quantum mechanical behaviour. At low temperatures quantum tunnelling is expected to be the dominant mechanism. At high temperatures, the transition is dominated by classical jumping over the barrier. In order to understand the process of H diffusion in Fe it is essential to study H trapping and migration over the whole range of temperatures covering both the quantum and classical dominated regimes and the cross over between them.

In this paper we apply quantum rate theory to calculate the transition rates as hydrogen and deuterium atoms jump from a vacancy in iron into a neighbouring metastable site. Knowledge of the transition rates and activation energies over a wide temperature range is essential for understanding the possible mechanisms of H diffusion and trapping in presence of vacancies. The precomputed transition rates along the minimal energy paths (MEP) between metastable sites can be employed as input data to a kinetic Monte Carlo (kMC) model of hydrogen diffusivity. The application of kMC for the study of H diffusion permits simulations in larger blocks of atoms for periods of time significantly longer than one can achieve with direct molecular dynamics simulation, which is essential for studying hydrogen migration and trapping in the presence of microstructural imperfections and consequent extraction of the diffusion coefficients.

## QUANTUM TRANSITION RATE

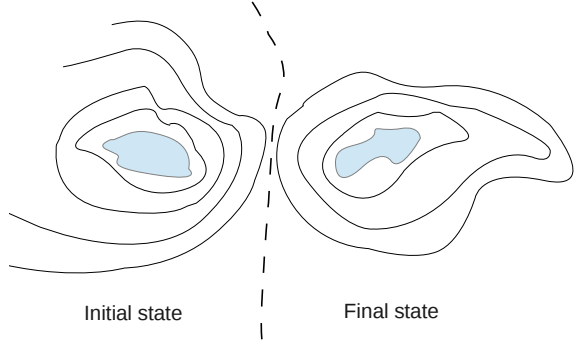

FIG. 1. A cartoon of a two dimensional potential energy surface with basins containing the reactants and products of a chemical reaction. In this case the initial and final states refer to a proton or deuteron trapped at certain interstitial sites in the bcc Fe lattice. These could be for example a bulk tetrahedral site and a trap site at a vacancy. The dotted line represents a choice of dividing surface, which in the two dimensional case is a line.

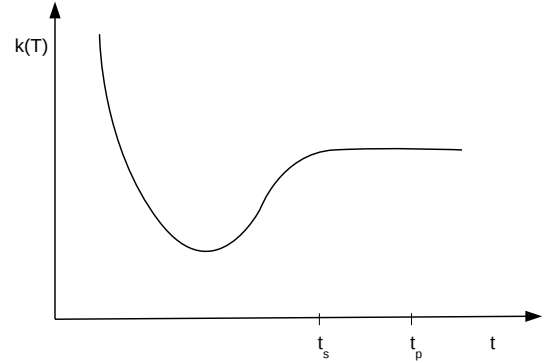

FIG. 2. A schematic plot of the characteristic behaviour of the reactive flux $k(T)$ as a function of time. In the initial stages of the chemical reaction the flux varies rapidly with time as the reactants cross and recross the dividing surface. It is this recrossing which is taken care of in the theory by the transmission coefficient and which is non uniform at times smaller than $t = t_s$. At a time $t = t_p$ one reaches the plateaux region and the reactive flux remains thereafter constant in time. This does not mean that recrossing no longer occurs, just that a steady state has been reached and an equilibrium constant can be defined [9].

In the transition state theory, we imagine a system of reactants in an initial state at time $t = 0$ able to cross an energy barrier into the final state as sketched in Fig. 1. This barrier acts as a saddle point in a complicated $N$-dimensional space of coordinates $q, \dots, q_N$. We can choose an $(N-1)$-dimensional "dividing surface" and imagine the system crossing this surface back and forth at the beginning of the reaction until it settles down into a steady state where the rates of crossing and recrossing are constant in time; this is sketched in Fig. 2. The chemical reaction rate is determined by correlating the flux through the dividing surface $s(q, \dots, q_N) = 0$ at time $t = 0$ with the projection onto the product side of the dividing surface at some later time $t = t_p$. This later time lies in the "plateau region" where all transient effects associated with the crossing and recrossing of the dividing surface have ceased. The exact quantum mechanical rate coefficient can be written as [9]

$$k(T) = \frac{\tilde{c}^{fs}(t_p)}{Q_r(T)}$$

where $Q_r(T)$ is the reactant partition function. The Kubo-transformed flux-side correlation function $\tilde{c}^{fs}$ can be written as

$$\tilde{c}^{fs}(t) = Tr\left[e^{-\beta H}\tilde{F}e^{iHt/\hbar}he^{-iHt/\hbar}\right]$$

Where $\beta = \frac{1}{kT}$ and $h$ is a projection operator onto the product side of the dividing surface

$$h(s) = \begin{cases} 1, & s > 0 \\ 0, & s < 0 \end{cases}$$

$\tilde{F}$ is the Kubo-transform of the flux operator $F = \frac{i}{\hbar}[H, h]$

$$\tilde{F} = \frac{1}{\beta} \int_0^\beta d\lambda\, e^{\lambda H} F e^{-\lambda H}$$

and $H$ is the Hamiltonian, or total energy function.

### Ring polymer rate theory

Ring polymer molecular dynamics (RPMD) [10] is an approximate method for calculation of Kubo-transformed correlation functions. This approach is based on the well-known isomorphism between the quantum statistical mechanics of distinguishable particles and the classical system of several fictitious particles connected by harmonic springs [11]. The resulting ring polymer molecular dynamics is governed by an effective Hamiltonian, which is derived from Feynman's path integral (PI). RPMD rate theory is essentially a classical reaction rate theory, within which the quantum transition rate is approximated by the transition rate of an $n$-bead harmonic ring polymer under an external potential $V(q_1, \dots, q_n)$. The rate constant is

$$k^{PI}(T) = \frac{\tilde{c}_{fs}^{PI}(t_p)}{Q_r^{PI}(T)}$$

where

$$\tilde{c}_{fs}^{PI}(t) = \frac{1}{2\pi\hbar} \int d^n\mathbf{p} \int d^n\mathbf{q}\, e^{-\beta_n H_n} \delta[\bar{s}(\mathbf{q})] \bar{v}_s(\mathbf{p}, \mathbf{q}) h[\bar{s}(\mathbf{q}_t)]$$

with $\beta_n = \frac{\beta}{n}$, $\bar{s}(\mathbf{q}) = s(\bar{q})$ and $\bar{v}_s(\mathbf{p},\mathbf{q}) = \frac{\partial \bar{s}(\mathbf{q})}{\partial \bar{q}} \frac{\bar{p}}{m}$, where $\bar{q} = \frac{1}{n}\sum_{i=1}^n q_i$ is the centroid of the ring polymer and $\bar{p} = \frac{1}{n}\sum_{i=1}^n p_i$ is its averaged momentum. The evolution of ring polymers is generated by an effective classical Hamiltonian

$$H_n(\mathbf{p}, \mathbf{q}) = \sum_{i=1}^n \frac{p_i^2}{2m} + V_n(\mathbf{q})$$

where

$$V_n(\mathbf{q}) = \sum_{i=1}^n \frac{1}{2} m\omega_n^2 (q_i - q_{i-1})^2 + V(q_1, \dots, q_n)$$

with $\omega_n = \frac{1}{\beta_n \hbar}$ and $q_0 = q_n$. The transition rate can be written as a product of two factors [12, 13]

$$k^{PI}(T) = \kappa(t_p) k^{QTST}(T) \tag{1}$$

where $k^{QTST}(T)$ is the quantum transition state theory (QTST) rate and

$$\kappa(t) = \frac{\langle \delta(q^+ - \bar{q})(\bar{p}/m)h[\bar{q}(t) - q^+]\rangle}{\langle \delta(q^+ - \bar{q})(\bar{p}/m)h(\bar{p})\rangle}$$

(2)

is a time-dependent transmission coefficient [13], where $q^+$ is the reaction coordinate at the dividing surface. The advantage of this factorisation is that it allows one to split the calculation of transition rate into purely statistical and dynamical parts. The transmission coefficient $\kappa(t)$ accounts for the recrossing of the transition state dividing surface and corrects the deficiencies in the choice of the dividing surface. It can be determined by evolving ring polymer trajectories of randomly selected centroid-constrained configurations of beads with centroids pinned at the dividing surface ($\bar{q} = q^+$). The ring polymer momenta are sampled from a Maxwell distribution contained in $e^{-\beta_n H_n}$. The resulting RPMD trajectories are evolved in time without a thermostat or dividing surface constraint to obtain $\bar{q}(t)$. The transmission coefficient is calculated by averaging over a large number of these trajectories in the numerator and denominator [13].

**QTST transition rate**

The QTST rate coefficient may be written

$$k^{QTST}(T) = \frac{1}{(2\pi\beta m)^{1/2}} \frac{Q^+(T)}{Q_r(T)} = \frac{1}{(2\pi\beta m)^{1/2}} \frac{\int d\mathbf{q}\, e^{-\beta V(\mathbf{q})} h(q^+ - \bar{q})}{\int d\mathbf{q}\, e^{-\beta V(\mathbf{q})} \delta(q^+ - \bar{q})}$$

(3)

where $Q_r(T)$ and $Q^+(T)$ are the reactant and transition state partition functions [14]. For proton transfer the partition function in reactant configuration in the reaction coordinate direction can be approximated to be of a Gaussian form with a width factor $(\beta m\omega^2)^{-1/2}$ [14]. This re-expresses the centroid density of the quantum system in terms of the centroid density of a reference action functional. Within this approximation, the QTST transition rate can be re-expressed as [14]

$$k^{QTST}(T) \simeq \omega \exp(-\beta \Delta F)$$

where $\omega$ is the frequency of oscillation of the proton in the $q$ direction. The activation energy

$$\Delta F = -k_B T \ln\left[\frac{Q^+(T)}{Q_r^+(T)}\right]$$

(4)

is the difference between free energies of transition state and reactant configuration with the centroid constrained on a surface parallel to the dividing surface in the vicinity of the reactant well [14]. $Q^+(T)/Q_r^+(T)$ is the Boltzmann factor in moving the reaction coordinate centroid variable from the reactant configuration to the transition state.

$k^{QTST}(T)$ can be calculated by performing thermodynamic integration based on the potential of mean force (PMF) profile along the reaction coordinate. In practice, one performs a series of constrained simulations with increasing values of $\bar{q}$ so as to obtain the mean force and calculates the free energy difference between transition and reactive states by numerical integration [12,15]. The sampling of the canonical distribution necessary for calculations of the mean force (MF) can be done by using RPMD in the presence of a thermostat. Our approach for calculation of the QTST rate coefficient is to use the Wang-Landau Monte Carlo (WLMC) algorithm [16] to calculate directly the partition functions participating in the QTST rate coefficient $k^{QTST}$ [17]. Unlike RPMD and Monte Carlo methods generating canonical distributions, the key idea of the WLMC method is to calculate the density of states directly by a random walk in energy space. This has the particular benefit that the partition functions and QTST transition rate can be calculated over a wide range of temperatures from a single simulation run. The path integral partition functions depend on two temperature independent functions

$$V_1(\mathbf{q}) = \sum_{i=1}^n \frac{m}{2\hbar^2}(q_i - q_{i-1})^2 \quad \text{and} \quad V_2(\mathbf{q}) = V(q_1, \dots, q_n)$$

$V_1(\mathbf{q})$ is related to the kinetic energy of the system and $V_2(\mathbf{q})$ accounts for the potential. Hence the density of states, which is an analogue of the classical density of states function, would depend on two variables, $v_1$ and $v_2$ [18]

$$\Omega(v_1, v_2) = \int d\mathbf{q}\,\delta(v_1 - V_1(\mathbf{q}))\,\delta(v_2 - V_2(\mathbf{q})) \tag{5}$$

With this density of states a partition function $Q(T)$ that describes the thermodynamics of the quantum system can be calculated for any temperature

$$Q(T) = \int_0^\infty dv_1 dv_2 \exp\left(-\frac{v_1}{\beta_n} - \beta_n v_2\right)\Omega(v_1, v_2) \tag{6}$$

The approach of Wang and Landau is to sample the density of states directly and, once known, calculate the partition function via (6).

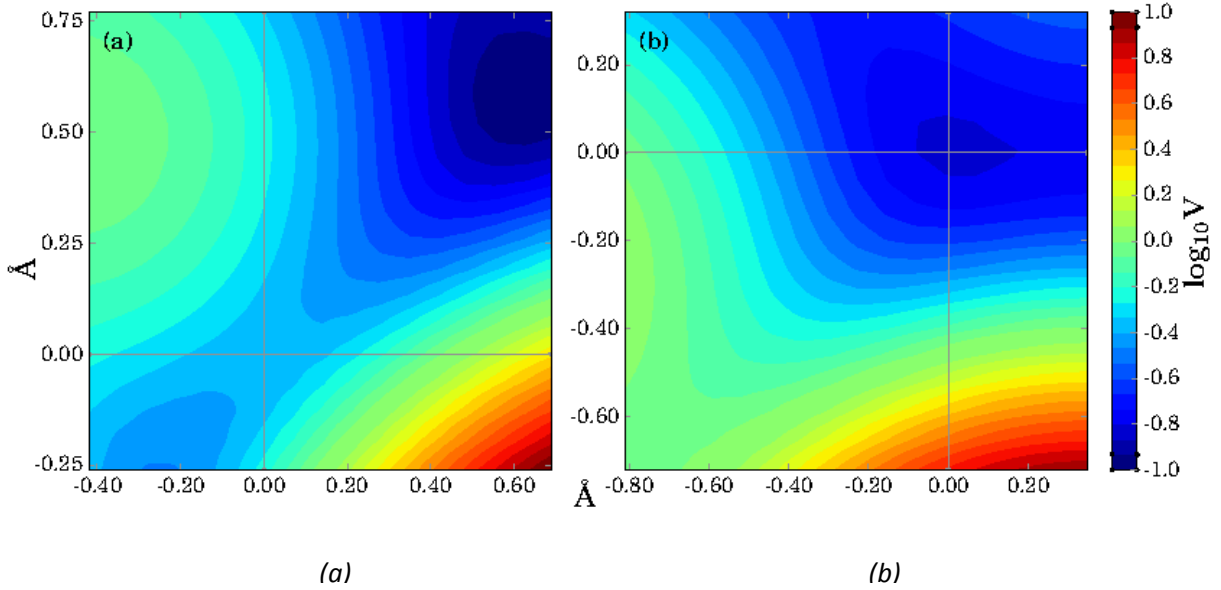

FIG. 3. Cross-sections of 3D potential energy surfaces for a H atom moving in the fixed atomic positions of the Fe atoms; (a) configuration in which the Fe atoms have been relaxed around a H atom in the trapping "reactant" site; (b) the Fe atoms are relaxed around a H atom held at the saddle point position between the trapping and neighbouring tetrahedral sites. The positions of the trap site and neighbouring tetrahedral sites are located at the origin of the coordinate system.

## RESULTS AND DISCUSSION

We have applied the approach described above to study the transition rates as hydrogen (H) and deuterium (D) atoms jump from a vacancy in iron into a neighbouring tetrahedral interstice. A crucial input for path integral calculations is the knowledge of the potential energy surface (PES). The interatomic forces and PES in magnetic iron, both pure and containing hydrogen impurities, in the present calculations, have been calculated by using a non-orthogonal self consistent tight binding model of the Fe–H system [19, 20].

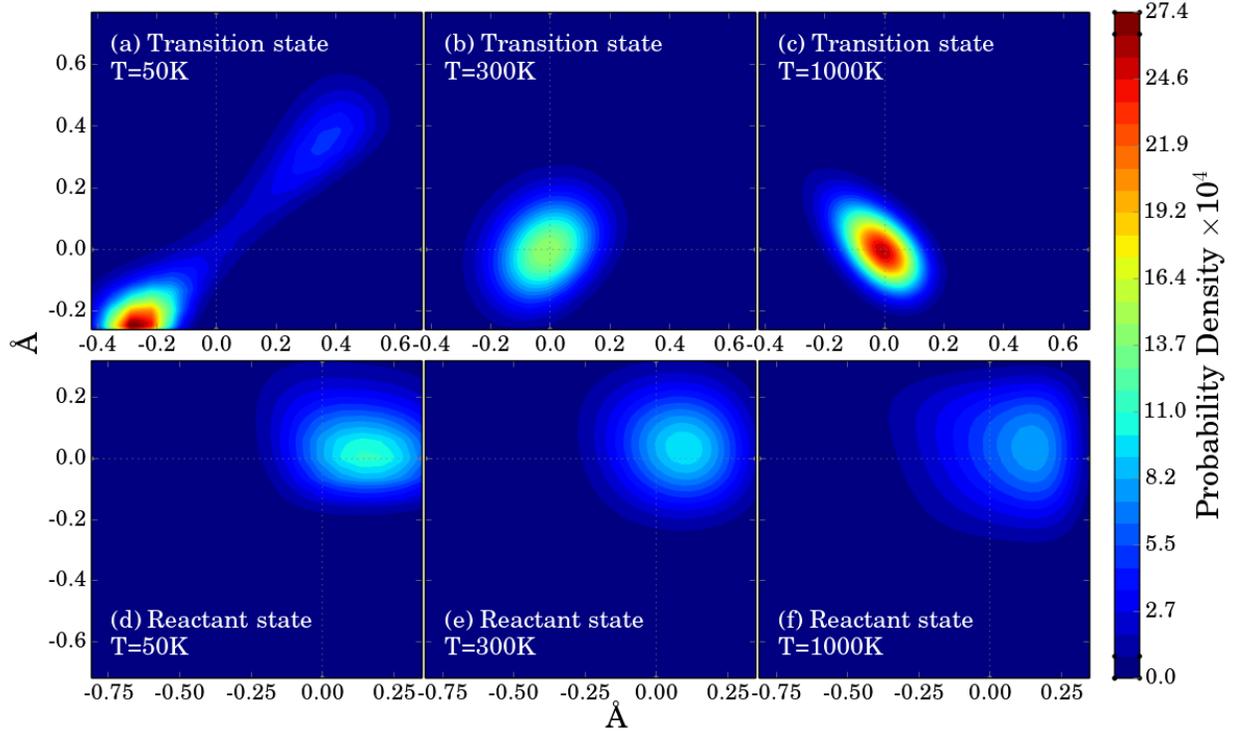

FIG. 4. Position probability density (PPD) of the hydrogen nucleus projected on a plane. Hydrogen is at transition state at temperature of (a) T =50K; (b) T =300K; (c) T =1000K; and at reactant state (d) T =50K; (e) T = 300K; (f) T =1000K. In each panel the saddle point (a)-(c) or the trapping site (d)-(f) are located at the origin of the coordinate system.

We constructed 3D potential energy surfaces (PESs) for the hydrogen motion in two fixed lattice configurations, corresponding to the fully relaxed lattice of 16 Fe atoms with a H atom in tetrahedral and saddle point configurations by performing a large set of TB total energy calculations corresponding to different positions of the hydrogen nucleus. In all these calculations the nuclei are treated as classical point particles. The calculated potential energy surfaces are shown in figure 3. Fig. 4 shows the 2D projection of the position probability density (PPD) of the H nucleus when its centroid is at the transition state (Fig. 4 a, b, c) and when it is confined in the potential well of the vacancy trapping site (Fig. 4 d, e, f). The same position probability densities in the case of the D atom are shown in Fig. 5. The figures show the probability distributions as functions of temperature. It is very clear that at low temperature the proton and deuteron in the transition state "split into two" with greatest PPD not at the saddle point but very close to the tetrahedral sites. The smaller part of the PPD is concentrated in the area of the vacancy trapping site. At low temperature the PPD of H and D nuclei in the reactant state contracts due to the freezing of the proton and deuteron into its lowest oscillator state. In the classical limit (high temperature) the H and D atoms can be considered as classical particles and their position probability densities are concentrated in the reactant well and in the vicinity of the dividing surface. Since D is the heavier, the confinement of its position probability to the dividing surface and reactant well is slightly more pronounced.

We have made WLMC simulations to calculate the free energy of H and D atoms at the trapping and transition states and hence to deduce the corresponding QTST barrier height (Eq. 4) as functions of temperature (Fig. 6a). The experimental values of the binding enthalpy for binding of deuterium atoms to vacancies in iron in the literature [21] is 0.63 eV for 1-2 D in a vacancy at an ambient temperature of 300 K. The present results are in good agreement with the experimentally observed data. At an ambient temperature of 300 K, the values calculated while taking quantum effects into consideration are 0.59 eV for D and 0.63 eV for H atoms. The corresponding QTST rates as functions of temperature calculated from (3) are shown in Fig. 6b. As should be expected, the quantum effects are more pronounced for H atoms. Fig. 6 shows that the calculated binding free energies for H and D at a vacancy increase when quantum effects are taken into account. The QTST transition rate is slightly higher in the case of deuterium. It is very significant that quantum effects cannot be ignored even above ambient temperature.

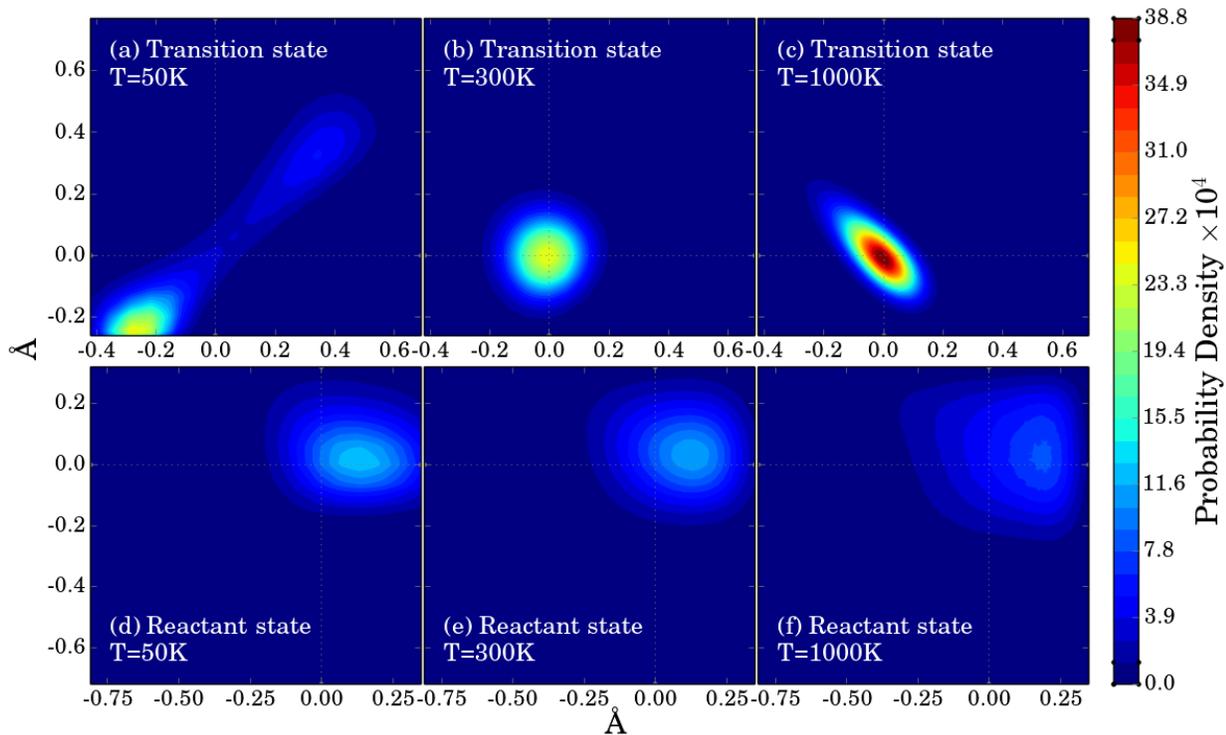

FIG. 5. Position probability density (PPD) of the deuterium nucleus projected on a plane. Deuterium is at transition state at temperature of (a) T =50K; (b) T =300K; (c) T =1000K; and at reactant state (d) T =50K; (e) T = 300K; (f) T =1000K. In each panel the saddle point or the trapping site are located at the center of the image. In comparison to figure 4 the particle is more confined even at ambient temperature. This reflects the heavier deuterium nucleus.

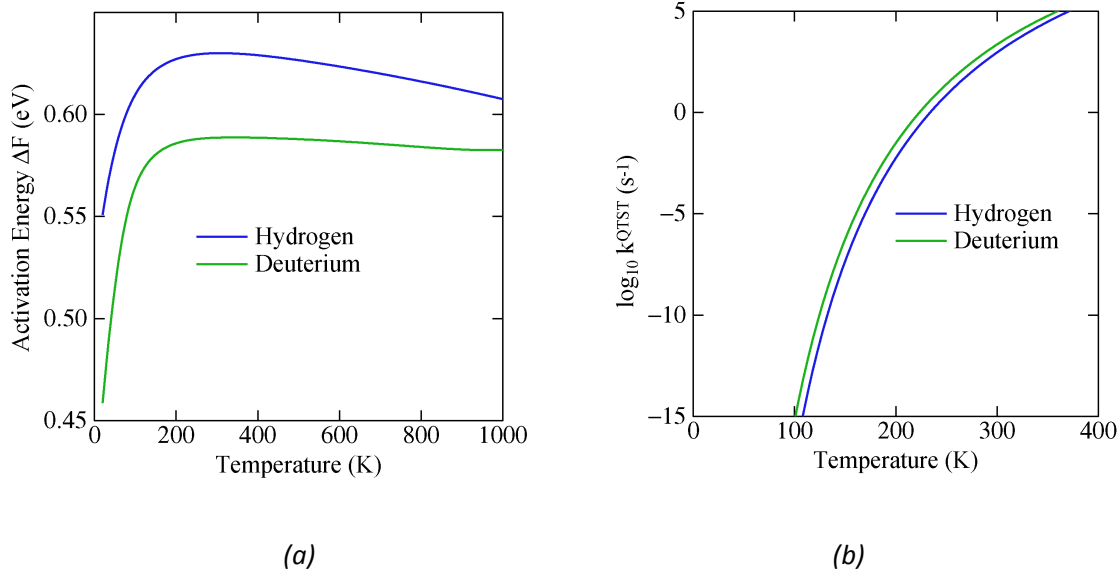

(a)  (b)

FIG. 6. *The free energy needed to carry the H and D atom from an initial stable position to a transition state (a) and the corresponding QTST rates (b) as functions of temperature.*

The final ingredient in the expression for the transition rate is the time-dependent transmission coefficient $\kappa(t)$ in Eq. (2). It was calculated at an ambient temperature of 300 K for proton and deuteron transfer by running $1.92 \times 10^7$ separate 1 ps RPMD trajectories from random starting configurations with the centroids confined to the dividing surface and momenta sampled from the Maxwell distribution. It was found necessary to make this number of trajectories in order to achieve a converged canonical sampling.

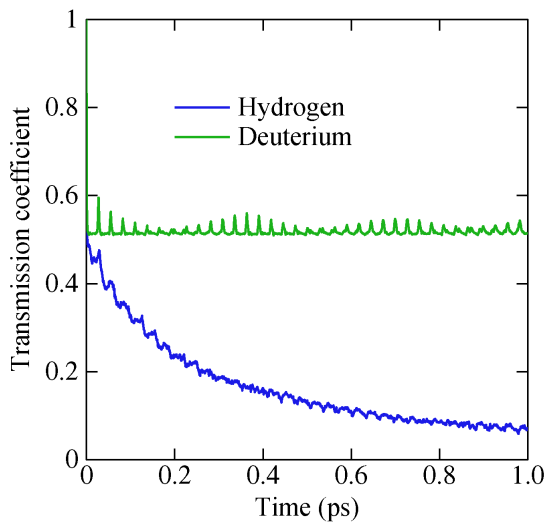

FIG. 7. *Transmission coefficient $\kappa(t)$ calculated at an ambient temperature of 300 K for proton and deuteron transfer.*

The resulting time-dependent transmission coefficients are shown in Fig. 7. It is clear from this figure that in both cases there is very significant recrossing of the transition state dividing surface. We observe significant differences between transmission coefficients calculated for proton and deuteron transfer. The transmission coefficient in the plateau region for proton transfer is found to be around 0.07 and that for deuteron transfer around 0.5. The separation of the rate constant into the QTST rate and a transmission coefficient is dependent on the choice of dividing surface (Fig. 1). This is straightforward if there is sufficient symmetry, for example in the case of bulk diffusion [17]. In the present instance of jumping in and out of a trap site from a bulk site the choice is arbitrary. Nevertheless because the transmission coefficient is significantly different from one (Fig. 7) it is clear that QTST overestimates the rate of proton transfer by more than an order of magnitude and the rate of deuteron transfer by a factor of about two. The reason for the overestimation is different for the two particles. The transmission coefficient in the case of the deuteron at room temperature coincides with the transmission coefficient calculated for a classical particle with the same mass. This fact leads to the conclusion that in this case the transmission coefficient is overestimated as a result of the choice of the dividing surface. In the case of the proton however the origin of the overestimation is the quantum effect due to the smaller mass of the proton. In this case the heavy Fe atoms impose a stronger kinematic constraint on the transmission of a hydrogen atom as a result of the quantum "delocalisation" of the proton. This very striking dependence on the mass of the diffusing particle clearly has implications in the interpretation of experiments which employ deuterium as a substitute for hydrogen, for example in order to produce unequivocal contrast in atom probe tomography [22]. Whereas static properties such as trap depths and sites will be reliably determined using deuterium, dynamical properties will be affected by our finding that the transition rate of hydrogen atoms jumping from a vacancy in iron into a neighbouring metastable site at ambient temperature is at least an order of magnitude lower than the transition rate in the case of deuterium atoms.

**CONCLUDING REMARKS**

We have applied quantum rate theory in its RPMD approximation to calculate the transition rates as hydrogen and deuterium atoms jump from a vacancy in iron into a neighbouring metastable site. Using the WLMC approach to calculate the QTST coefficient we determine transition rates and corresponding activation energies over a wide range of temperatures covering both the quantum and classical dominated regimes. We find that quantum effects lead to an increase of the transition rate activation energy and to very significant decrease of the transmission coefficient. These effects are significant even at ambient temperature. As a result of the light mass of the H nucleus, our quantum transition rate calculations indicate that hydrogen atoms are more deeply trapped at a vacancy than is predicted by density functional theory results.

We conclude by noting an important difference between our PI WLMC approach, path-integral Monte Carlo and path-integral molecular dynamics algorithms. The joint density of states of two or more variables (5) is useful because quantities such as the free energy can be calculated as a function of temperature from a single simulation run. The extra information does not come for free and for large systems it becomes expensive to converge Wang-Landau sampling. However, this difficulty can largely be eliminated by the much greater parallelisation of path-integral WLMC than MD or MC methods. In addition, an immediate disadvantage of the thermodynamic integration approaches using conventional canonical MC or MD is that it is necessary to perform many simulations of a system at physically uninteresting intermediate values, where the potential is unphysical. Only initial and final configurations correspond to actual physical states. Nevertheless, the intermediate averages must be accurately calculated in order for the integration to yield a correct result. Instead, each WLMC move samples an actual physical state which contributes to the convergence of Monte-Carlo procedure. Thus, PI WLMC becomes a competitive method with efficiency that rivals path-integral molecular dynamics and Monte-Carlo approaches. The computationally less expensive kinetic Monte-Carlo method (kMC), using quantum transition rates calculated by the PI WLMC technique proposed in this article, opens the way to studying the dynamics of hydrogen migration and trapping in presence of vacancies over a wide temperature range.

**ACKNOWLEDGMENTS**

We are grateful to the European Commission for funding under the Seventh Framework Programme, grant number 263335, MultiHy (multiscale modelling of hydrogen embrittlement in crystalline materials). We thank Dimitar Pashov, Matous Mrovec and Nick Winzer for assistance and valuable conversations.

[1] K. Kiuchi and R. B. McLellan, Acta Metall. 31, 961 (1983).

[2] H. Hagi and Y. Hayashi, Trans. Jpn. Inst. Met. 28, 368 (1987).

[3] Y. Hayashi, H. Hagi, and A. Tahara, Z. Phys. Chem. (Neue Folge) 164, 815 (1989).

[4] M. Nagano, Y. Hayashi, N. Ohtani, M. Isshiki, and K. Igaki, Scr. Met. 16, 973 (1982)

[5] Y. Katz, N. Tymiak, and W. W. Gerberich, Eng. Fract. Mech. 68, 619 (2001).

[6] R. Kirchheim, Prog. Mat. Sci. 32, 261 (1988).

[7] R. Kirchheim, Acta Met. 35, 271 (1987).

[8] A. Kumnick and H. Johnson, Acta Met. 28, 33 (1980).

[9] W. H. Miller, S. D. Schwartz, and J. W. Tromp, J. Chem. Phys. 79, 4889, (1983).

[10] I. Craig and D. Manolopoulos, J. Chem. Phys. 121, 3368 , (2004).

[11] R. P. Feynman, Statistical Mechanics—a set of lectures (Benjamin, Reading, Mass., 1972).

[12] D. Chandler, J. Chem. Phys. 68, 2959, (1978).

[13] R. Collepardo-Guevara, I. Craig, and D. Manolopoulos, J. Chem. Phys. 128, 144502, 2008.

[14] G. A. Voth, J. Phys. Chem. 97, 8365 (1993).

[15] M. E. Tuckerman, Statistical Mechanics: Theory and Molecular Simulation, (OUP, 2010).

[16] F. Wang and D. P. Landau, Phys. Rev. Lett. 86, 2050 (2001).

[17] I.H. Katzarov, D. Pashov and A. T. Paxton, Phys. Rev. B 88, 054107, (2013).

[18] P. N. Vorontsov-Velyaminov and A. P. Lyubartsev, J. Phys. A: Math. Gen. 36, 685 (2003).

[19] A. T. Paxton and M. W. Finnis, Phys. Rev. B 77, 024428 (2008).

[20] A.T. Paxton, C. Elsässer, Phys. Rev. B 82, 235125, (2010).

[21] F. Besenbacher, S. M. Myers, P. Nordlander, J. K. Norskov, J. Appl. Phys. 61 (5), 1788 (1987).

[22] J. Takahashi, K. Kawakami and Y. Kobayashi, T. Tarui, Scr. Mat., 63, 261 (2010).